 \documentclass[authoryear,preprint,review,12pt]{elsarticle}

\usepackage{amssymb}

\journal{New Astronomy}

\begin{document}

\begin{frontmatter}



\title{Light Curve Analysis of Hipparcos Data for the Massive O-type
Eclipsing Binary UW CMa}


\author{E. A. Antokhina$^1$, M. Srinivasa Rao$^2$\corref{corl} M. Parthasarathy $^3$}

\address{$^1$ Sternberg Astronomical Institute, Universitetskij pr. 13, Moscow 119992, Russia, email: elant@sai.msu.ru\\
$^2$  Indian Institute of Astrophysics, Koramangala, Bangalore 560034,  India: email: msrao@iiap.res.in \\
$^3$  Aryabhatta Research Institute of Observational Sciences, Nainital 263129, India, email: m-partha@hotmail.com 
}
 \cortext[cor1]{Tel/Fax: 91 80 25530672/91 80 25534043}

\begin{abstract}

Hipparcos photometric data for the massive O-type binary UW CMa were
analysed within the framework of the Roche model. Photometric
solutions were obtained for five mass ratios in the 
$q=M_2/M_1=0.5-1.5$ range. The system is found to be in a contact
configuration.  Independently of $q$, the best-fitting model solutions
correspond to the orbital inclination $i \sim 71^\circ$ and the temperature of the
secondary component $T_2 \sim 33500\,K$, at the fixed temperature of the primary
$T_1=33750 K$.  Considering that the spectrum of the secondary is very
weak, photometric solutions corresponding to the contact configuration
favor the mass ratio $q$ smaller than unity (in which case the
luminosity of the secondary is smaller than that of the primary). The
absolute parameters of the system are estimated for different values of the
mass ratio.

\end{abstract}


\end{frontmatter}


\label{}



\section{Introduction}

The spectroscopic eclipsing binary UW CMa (=29 CMa= HD57060) is an
interesting massive O-type close binary system (P=4.4 days).  
Light curves of UW CMa have been analysed by various authors. Since the
spectrum of the secondary component is extremely weak and hardly
detected, the spectroscopic mass ratio is uncertain. For this reason,
different assumptions on the mass ratio in the system were made in
various photometric studies.
 
Parthasarathy (1978) carried out an analysis of BV observations of Doss
(1967) by the method of Russell and Merrill (1952). He determined absolute
dimensions of UW CMa by combining derived photometric elements and 
spectroscopic data by Struve et al. (1958), which suggested that the mass
ratio is $q=M_2/M_1=1.20$ (secondary more massive).

Leung and Schneider (1978a) and Bagnuolo et al. (1994) carried out a
detailed analysis of spectroscopic and photometric data on UW CMa, but their
results are not quite consistent. Leung and Schneider (1978a) analysed the
photographic light curve of Seyfert (1941) and BV light curves of Ogata and
Hukusaku (1977) using the Wilson and Devinney (1971) algorithm. The authors
searched for solutions for five different fixed values of the mass
ratio in the $q=0.75-1.30$ range. As they noticed earlier for
several other systems, a unique value for the mass ratio usually could not
be determined from a light curve alone (Leung and Schneider, 1978b). Involving
some arguments related to the luminosity ratio in the system, the authors
came to a conclusion that the most likely value of the mass ratio was $q \le 1$
(primary more massive). The temperature of the O7f star was fixed at
$T_1=43000 \; K$, and the corresponding temperature of the secondary $T_2
\sim 40000 \;K$ was obtained. The authors reported five sets of photometric
parameters and concluded that a contact configuration of this binary was
fairly certain.
 
Bagnuolo et al. (1994) carried out a comprehensive analysis of the UW CMa UV
spectra from the IUE archive. The tomography algorithm was used to separate
spectra of the two stars. The authors found the mass ratio
 $q= 1.2$ using three independent methods: (i) fitting
cross-correlation functions; (ii) comparing radial velocity semi-amplitude
$K_1$ and $V  {\rm sin} i$ (Gies and Bolton, 1986); (iii) measuring the
goodness-of-fit of shifted secondary spectra produced by the tomography
algorithm for an assumed grid of mass ratios. They also obtained a new
spectral classification for the primary (O7.5-8 Iabf) and the secondary
(O9.7 Ib). The intensity flux ratio of the stars in the UV was found to be
$r=0.36 \pm 0.07$ (primary brighter). By using spectral type calibration of
Howart and Prinja (1989) new temperatures of stars were estimated:
$T_1=33750 K$  and $T_2=29000 K$.

Then Bagnuolo et al.
fitted the V-band light curve (van Genderen et al., 1988) and the UV light curve
(Eaton, 1978) using GENSYN code (Mochnacki and Doughty, 1972; Gies and
Bolton, 1986). It was shown that a good fit could be obtained for a
semi-contact configuration, the orbital inclination  $i=74^\circ \pm 2^\circ$ and
a reasonable intensity ratio $r < 0.5$. The photometric model implies that the
primary fills its inner Roche lobe (fill-out ratio $f_1=1.0$), while the
fill-out ratio of the secondary is $f_2=0.7$. The authors conclude that the
radius of the secondary is about $70-80 \%$ as large as that of the primary.
This result is inconsistent with the model of Leung and Schneider (1978),
which suggested a large radius of the secondary.

\section{Hipparcos light curve of UW CMa}

New photometric data on UW CMa (HIP 35412) were obtained during the
Hipparcos mission. For our light curve analysis we used photometric
observations in the broad-band Hp system (effective wavelength $\lambda_{eff}
\sim 4500$ \AA). The light curve includes 217 data points obtained between
1990 March 31 and 1993 March 3. Orbital phases were calculated with the
following ephemeris for the primary minimum (Herczeg et al., 1981):

$$
JD(hel) \; Min I = 2440877.563 + 4^d.39336 \cdot E
$$

\noindent
The primary minimum is due to the eclipse of the more luminous star by the less 
luminous companion. The asymmetries in the light curve noticed by the earlier 
observers are also present in the Hipparcos light curve. 
The observed light curve (Figs.1-5) shows almost symmetrical primary minimum
while the secondary minimum is asymmetrical. Both branches of the secondary
minimum are not smooth, notably the ascending branch is more distorted.
These distortions are likely due to a mass flow, gas streams and/or stellar
wind in the binary. 

\section{Analysis}

In the current study  we adopted an approach similar to that of Leung and
Schneider (1978). We have analysed the Hippparcos photometric light curve
in two fashions: (i) including all observations; (ii) omitting orbital phases
$0.5-0.7$ on the ascending branch of the secondary minimum (the most distorted
part of the light curve). It turned out that in both cases the obtained
solutions were nearly identical. For this reason, significance levels of the
derived model parameters were estimated for the second case only.
   
The Hipparcos photometric light curve was analysed within the framework of the
Roche model in eccentric orbit, similar to Wilson's (1979) model. The
algorithm is described in detail by Antokhina (1988, 1996). Here we
describe its main features only briefly.
The computer code allows one to calculate a radial velocity curves,
monochromatic light curves and absorption line profiles of the stars
simultaneously, either for a circular or an eccentric orbit.
Axial rotation of the components may be non-synchronized with the
orbital revolution. Tidal distortion of the components as well as their
mutual radiative heating are taken into account. The intensity of the
radiation coming from an elementary area of the stellar surface and its
angular dependence are determined by the temperature of the star,
gravitational darkening, limb darkening, and heating by radiation from the
companion. Input parameters of the model are summarized in Table 1.


 \begin{table}[h!]
\centering
 \caption{Input Parameters of the synthesis program}
 \label{Input_Par}

\medskip

 \begin{tabular}{ll}
 \hline
 \noalign{\smallskip}
 Parameters & Description  \\
 \noalign{\smallskip}
 \hline
 \noalign{\smallskip}

\hbox to 25mm{$q=M_2/M_1$ \dotfill}      & Mass ratio \\
\hbox to 25mm{$e$ \dotfill}              & Eccentricity \\
\hbox to 25mm{$\omega$ \dotfill}         & Longitude of periastron, star No.1 \\
\hbox to 25mm{$i$ \dotfill}              & Orbital inclination \\
\hbox to 25mm{$\mu_1$, $\mu_2$ \dotfill} & Roche lobe filling coefficients, $\mu=R/R^*$, where $R$ is the \\
                                         & polar radius of a star and $R^*$ is the polar radius of the \\
                                         & corresponding inner critical Roche lobe at periastron position \\
\hbox to 25mm{$T_1$,$T_2$ \dotfill}      & Average effective temperatures of the components \\
\hbox to 25mm{$\beta_1$,$\beta_2$ \dotfill} & Gravity darkening coefficients (the temperature of an \\
                                         & elementary surface area
                                           $T=T_{1,2}\times{({g\over{<g>_{1,2}}})}^{\beta_{1,2}}$, where \\
                                         & $g$ and $<g>$ are the local and mean gravity
                                           acceleration) \\
\hbox to 25mm{$A_1$,$A_2$ \dotfill}      & Bolometric albedos (coefficients of reprocessing \\
                                         & of the emission of a companion by ''reflection'') \\
\hbox to 25mm{$F_1, F_2$ \dotfill}       & Ratio of surface rotation rate to synchronous rate \\
\hbox to 25mm{$x_{1,2}$, $y_{1,2}$ \dotfill}     & Limb darkening coefficients (see the text) \\
\hbox to 25mm{$l_3$ \dotfill}            & Third light \\
\hbox to 25mm{$\lambda(n)$ \dotfill}     & Effective wavelengths of monochromatic light curves \\

\noalign{\smallskip}
\hline
\end{tabular}
\end{table}

\subsection {Input Parameters}

We fixed some parameters which values were defined in previous studies of
the system or can be assumed from global stellar properties. A light curve
solution is only sensitive to the temperature difference between the stars,
thus the temperature of one star has to be fixed. Usually it is the
temperature of the primary, which can be determined more reliably. As
we mentioned earlier, Bagnuolo et al. (1994) obtained new spectral
classification of the primary (O7.5-8 Iabf) and the secondary (O9.7 Ib) and
their temperatures $T_1=33750 K$ and $T_2=29000 K$ using the spectral type
calibration from Howart and Prinja (1989). Using these data we fixed the
average effective temperature of the primary star at $T_1=33750$~K.

We fixed the gravity-darkening coefficients $\beta_1=\beta_2=0.25$ and
albedos $A_1=A_2=1$ to values typical for early type stars. A non-linear
``square-root'' limb darkening law (Diaz-Cordoves and Gimenez, 1992;
Diaz-Cordoves et al.,1995; Van Hamme,1993) was used:

$$
I(\cos\gamma)=I(1)[1-x(1-\cos\gamma)-y(1-\sqrt{\cos\gamma})],
$$

\noindent where $\gamma$ is the angle between the line of sight and the normal to the
surface, $I(1)$ is the intensity at $\gamma = 0$, and $x,y$ are limb
darkening coefficients. As shown by Van Hamme (1993), this is the most
appropriate limb--darkening law at optical wavelengths for $T\geq10000$~K.

The Hipparcos light curve indicates circular orbit, so we fixed eccentricity
at $e=0$. The rotation of both stars was assumed to be synchronous with the
orbital revolution $F_1=F_2=1$.

\subsection {Adjustable  Parameters}

Thus the adjustable parameters of the models  were (i) the Roche lobe filling
coefficients for the primary and the secondary $\mu_1,\mu_2$ (these parameters
define the dimensionless surface potentials $\Omega_1, \Omega_2$), (ii) the
average effective temperature of the secondary star $T_2$, and (iii) the
orbital inclination $i$. We fitted the model light curves for five fixed
values of mass ratios (see below).

The search for the adjustable parameters was done with the well-known Simplex
algorithm (Nelder and Mead's method) (Himmelblau, 1971; Kallrath and
Linnell, 1987). In the vicinity of the minima found, additional calculations
were done on a fine grid, to explore the details of the  shape of the residuals
surface and to determine the confidence intervals for the free parameters of
the model. The confidence intervals for the parameters were estimated using
$\chi^2$ test at a confidence level of $1\%$.

Since the mass ratio is unknown we searched for solutions at its five
different values, $q=0.5, 0.75, 1.0, 1.25, 1.5$. The
resulting parameters for the five solutions are presented in Table 2. The
corresponding model light curves along with the observed light curve are
shown in Figs.1-5.
The sky plane view of UW CMa for one of the solutions from Table 2 is
shown in Fig 6.

\begin{figure}
\centering
\includegraphics [width=15cm] {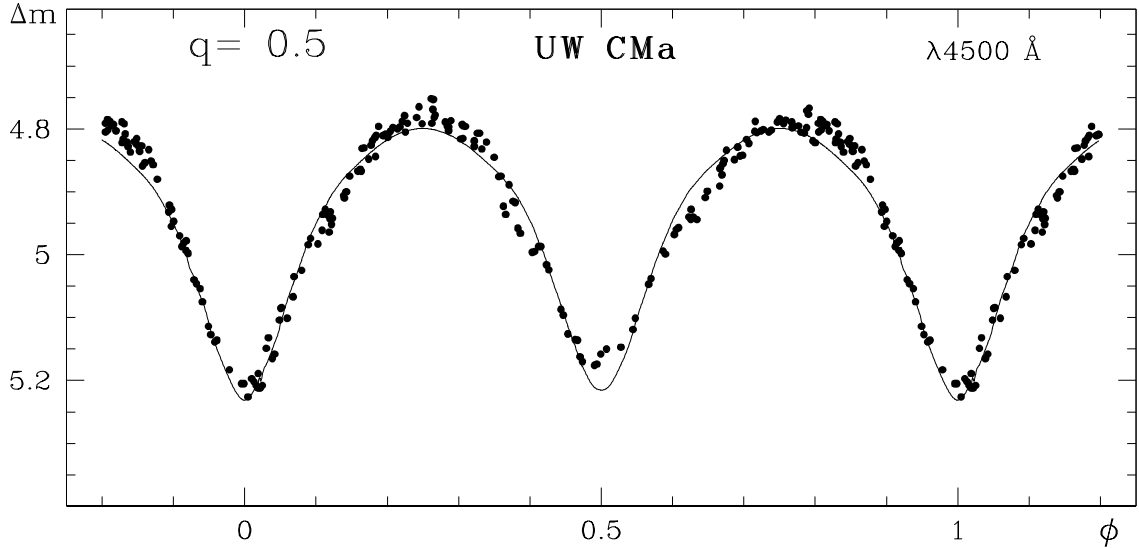}
\caption{Observed and model light curves for
$q=0.5$. For this mass ratio minimal deviation
exceeds the critical value $\chi2$ at the confidence
level of $1\%$}
\end{figure}

\begin{figure}
\centering \includegraphics[width=15cm]{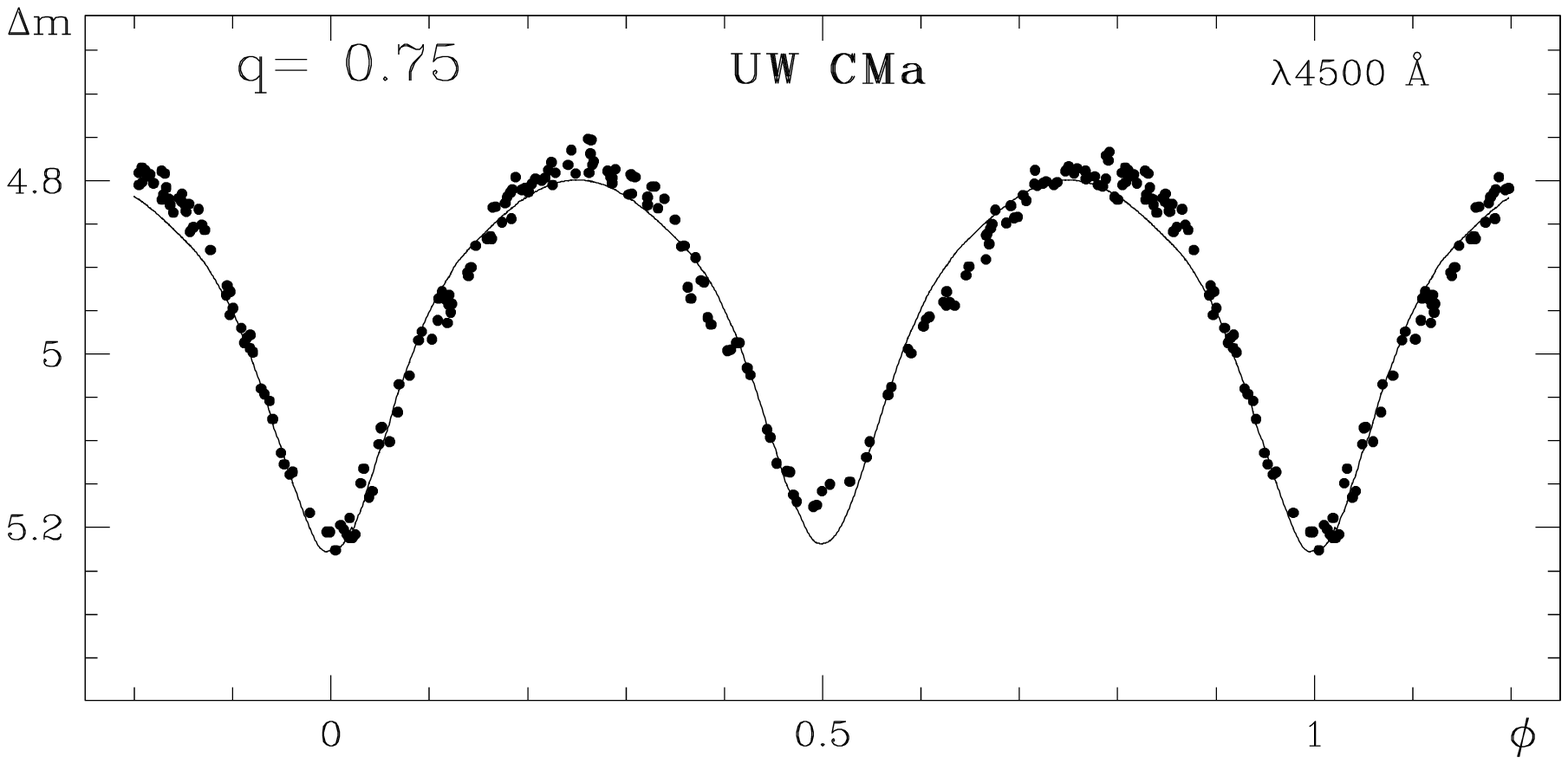}
\caption{Observed and model light curves at $q=0.75$}
\end{figure}

\begin{figure}
\centering \includegraphics[width=15cm]{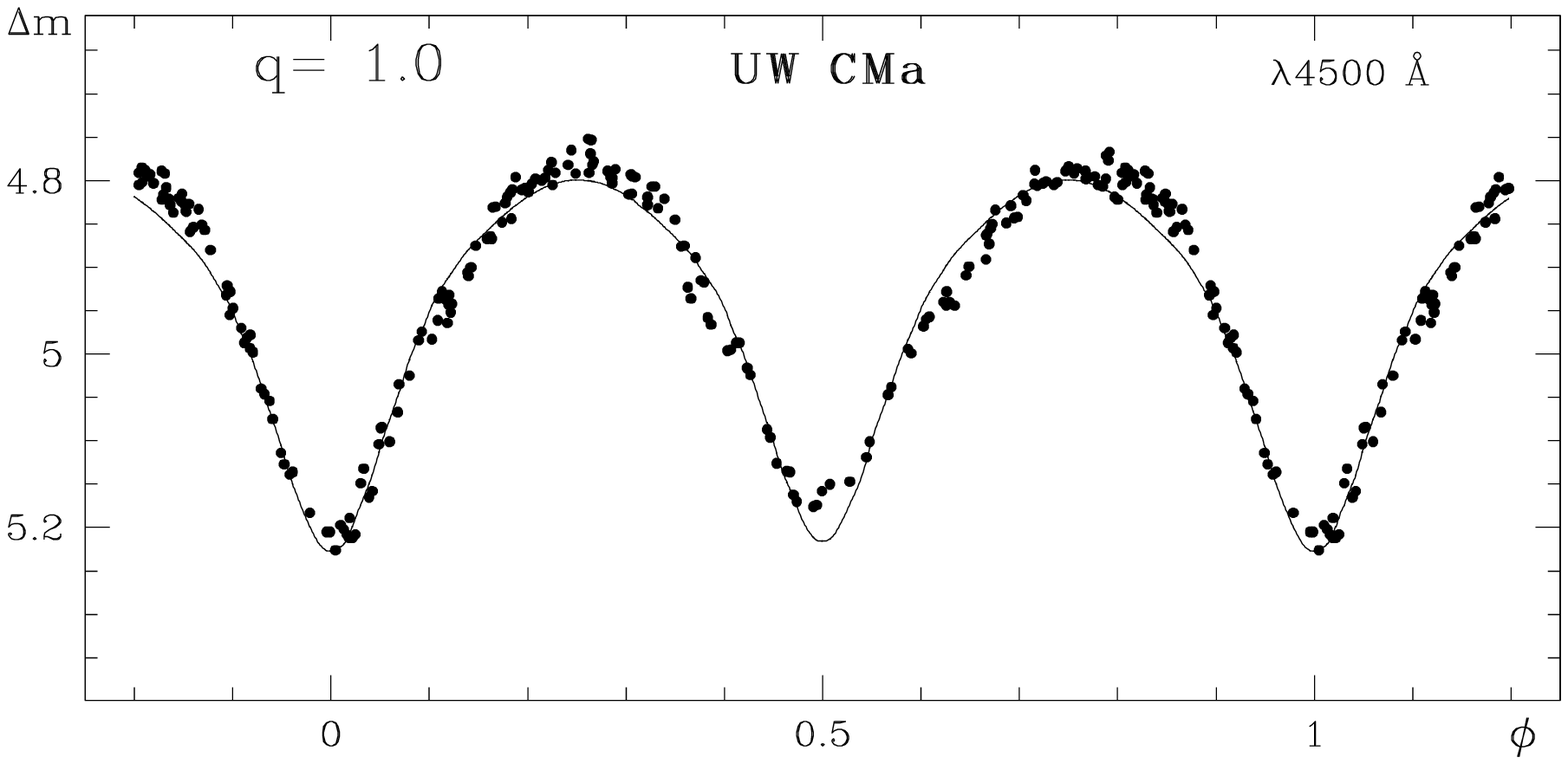}
\caption{Observed and model light curves at $q=1.0$}
\end{figure}

\begin{figure}
\centering \includegraphics[width=15cm]{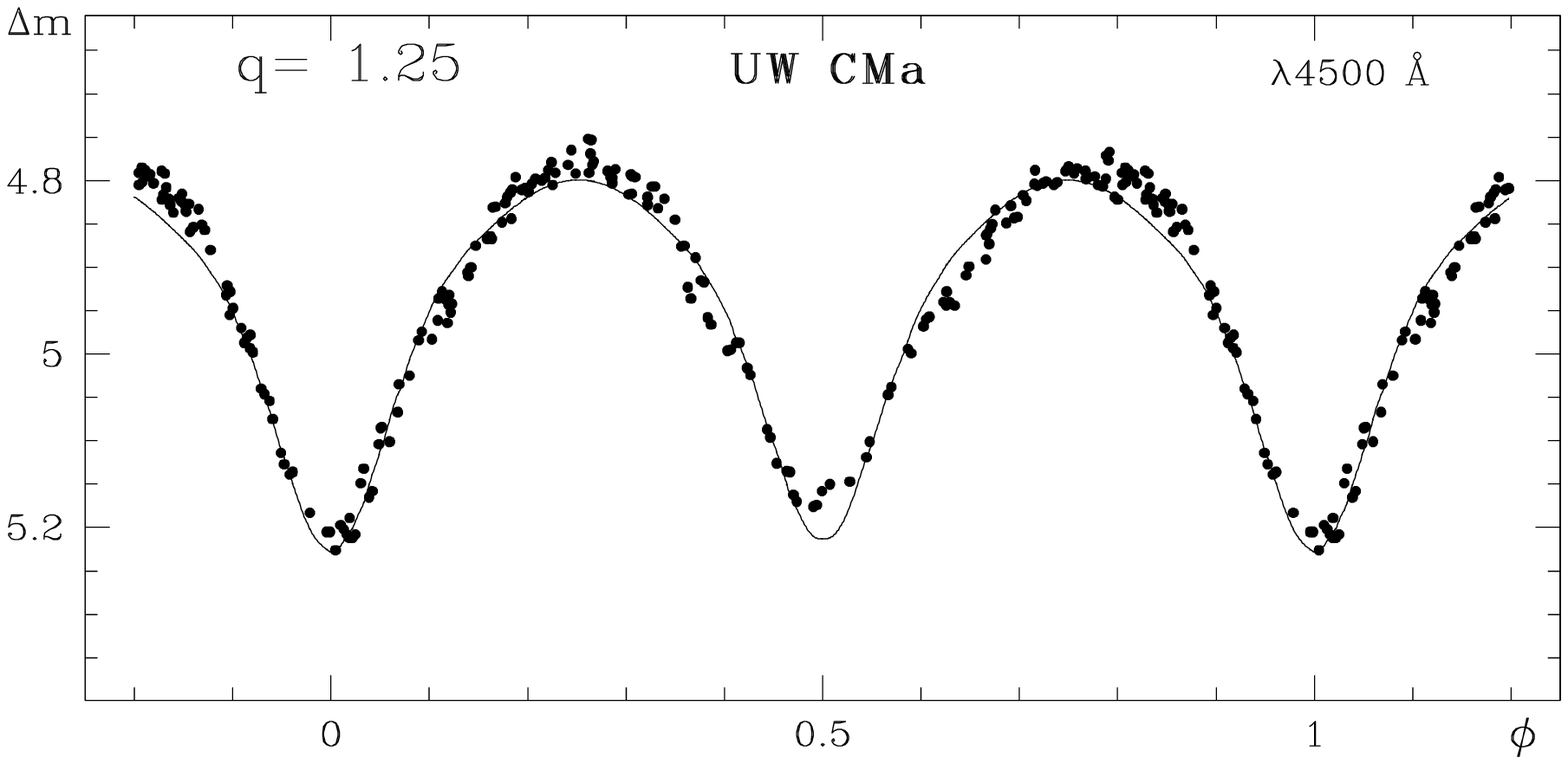}
\caption{Observed and model light curves at $q=1.25$}
\end{figure}

\begin{figure}
\centering \includegraphics[width=15cm]{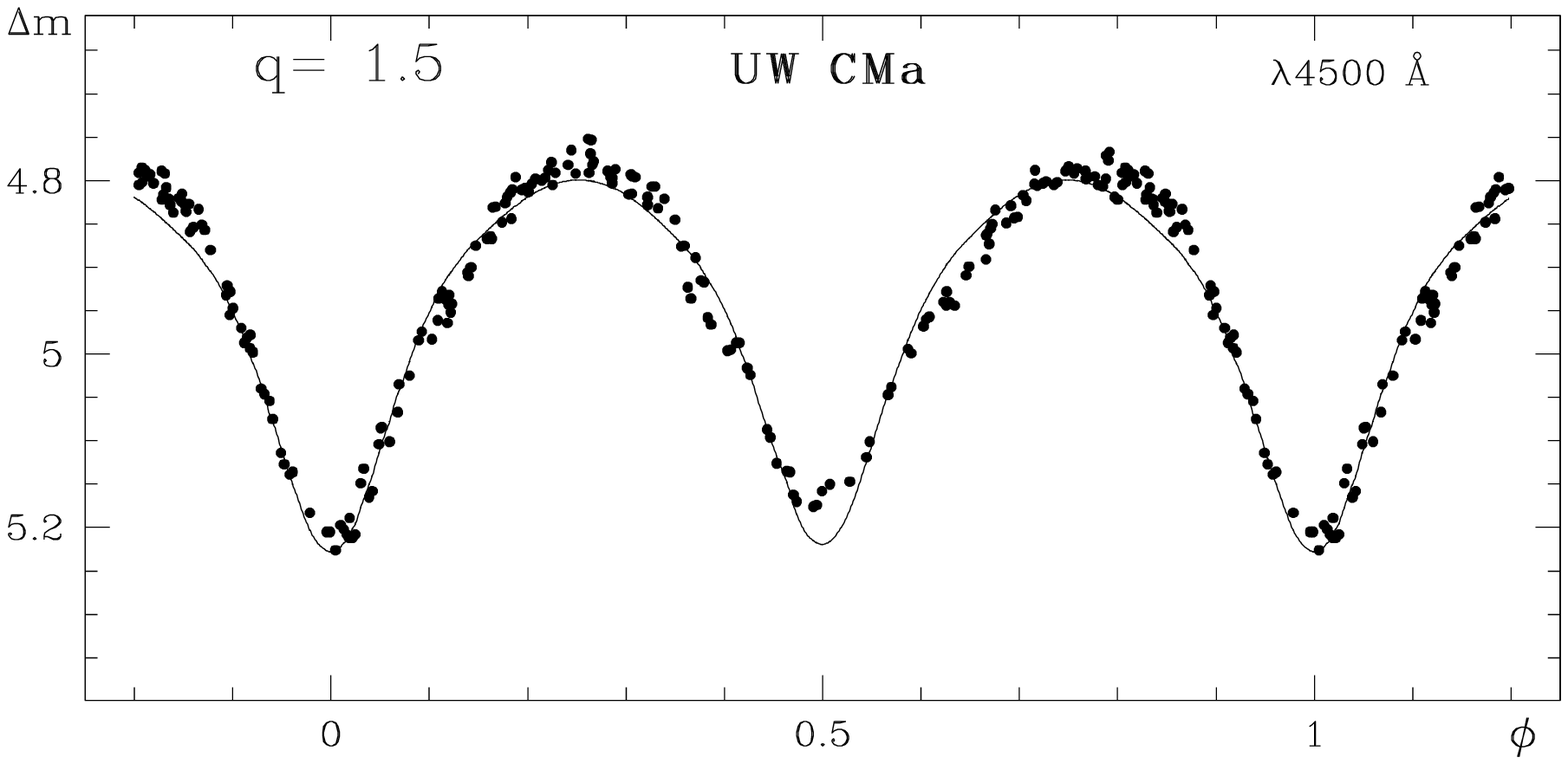}
\caption{Observed and model light curves at $q=1.5$}
\end{figure}

\begin{figure}
\centering \includegraphics[width=16cm]{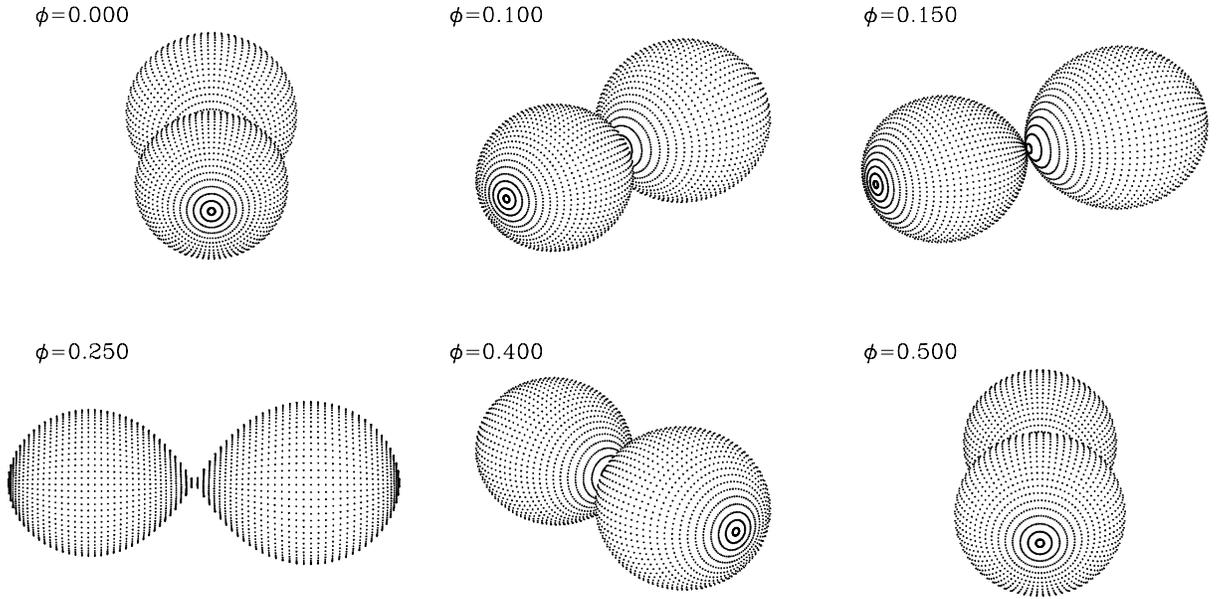}
\caption{
The sky plane view of UW CMa at different orbital phases.
The mass ratio $q=0.75$}
\end{figure}

\newpage

\subsection {Absolute Dimensions} 

From Table 2 it can be concluded that for all five mass
ratios the obtained solutions are rather close. Indeed, the $\chi^2$ values
are rather similar except the model for $q=0.5$ where $\chi^2$ is somewhat
larger. The values of the adjustable parameters $\mu_1,\mu_2$, $T_2$, $i$
are also rather close for all models with $q>0.5$. The absolute dimensions
of UW CMa for different values of $q$ are given in Table 3. While computing
the absolute values we used the semi-amplitude of the radial velocity curve
of the primary component $K_1=224.5 \; km/s$ (Stickland, 1989).  

\section {Results and Discussion} 

Current analysis of the Hipparcos light curve of UW CMa allows us to
definitely state that the system is in a contact configuration. This result
confirms a similar conclusion made by Leung and Schneider (1978a) from their
analysis of the light curves by Seyfert (1941) and Ogata \& Hukusaku (1977).
Another conclusion similar to that of Leung and Schneider (1978a), is that in
a contact configuration the mass ratio $q$ should be smaller than
unity. Indeed, the spectrum of the secondary is very weak which implies
that $L_2/L_1$ (visual) is less than unity.

These conclusions are in disagreement with those reached by Bagnuolo et al.
(1994) from the analysis of V (van Genderen et al., 1988) and UV (Eaton,
1978) light curves. The authors argue for a semi contact configuration
(the secondary underfills its Roche lobe) to explain the low luminosity of the
secondary. However, from the appearance of their light curves
(Bagnuolo et al., 1994) it seems they could be better fitted in a contact
model than in a semi contact one.

A more serious problem concerns the mass ratio in the system. Indeed,
the contact configuration favors a small mass ratio, $q < 1$.
However, Bagnuolo et al. (1994) obtained an estimate $q= 1.2$ from the UV data.
Their estimate was consistent for three different methods they used. Presently,
it is difficult to resolve the issue and to make a final conclusion on the mass
ratio in the binary. 
Detailed spectroscopic study could probably allow one to derive the mass ratio.
However such a study is beyond the scope of the present paper.

The following definitive conclusions can be drawn from our analysis:
 
\begin{enumerate}

\item The system is in contact configuration.

\item Independently of a particular value of $q$, the best-fitting model
solutions correspond to the following values of the free parameters:
$i=71^\circ.0 - 71^\circ.6$ and $T_2=33300 - 33700\,K$.

\item It is impossible to reliably determine the value of $q$ from the
light curve solution alone. This conclusion was predictable and has been
already discussed by Leung and Schneider (1978a,b).

\end{enumerate}

One possibile way to explain the discrepancy in the determination of the
mass ratio is a possibility that the geometry of UW CMa is different
from the standard Roche geometry. The secondary component of a contact
system could be a star surrounded by an optically and geometrically thick
envelope (disk). Such a model was used by Antokhina and Cherepashchuk
(1987) and Antokhina and Kumsiashvili (1999) in the light curve analysis of
the massive interactive binary system RY Sct. In this model a binary
consists of a primary component treated as a normal star in a Roche model
and a disk-shaped secondary (oblate spheroid). This model was first
suggested by Wilson (1974) for the analysis of $\beta$ Lyrae.

Another model where the secondary star is surrounded by a disk was used by
Djura\v{s}evi\'c et al. for light curve analysis of RY Sct (Djura\v{s}evi\'c et
al., 2008) and V448 Cyg (Djura\v{s}evi\'c et al., 2009). RY Sct is an active
mass-transferring system. It contains the O9.5-B0 primary filling its Roche
lobe which transfers mass to the more massive and hotter (although
apparently fainter) secondary component, hidden within a dense accretion
disk (Giuricin and Mardirossian, 1981; Antokhina and Cherepashchuk, 1987,
Djura\v{s}evi\'c et al., 2008; Grindstrom et al., 2007; and references therein).
Giuricin and Mardirossian (1981) presented a list of OB-binaries, which
could presumably be at the same evolutionary stage as RY Sct. They included
UW CMa in this list. This again could suggest existence of a dense accretion
disk in the system.

In the current paper we do not apply a disk model to light curve analysis of
UW CMa. The considerations above are just speculations on one possible way
to resolve the mass ratio problem in the system. They are not arguments for
the disk existence in UW CMa. To apply a disk model, we need some
observational evidences for the presence of a disk. Such arguments (if any)
could possibly be obtained from spectroscopy.

UW CMa is a massive early type contact binary. Spectroscopic
observations in UV, optical, and X-ray domains reveal colliding winds.
The system also shows active mass transfer and mass loss. The evolutionary
time scale of massive early type contact systems is relatively
short. UW CMa appears to be close to the common envelope phase of its
evolution. The common envelope evolution in massive close binary stars leads
to various degrees of stripping the envelope  of the more massive star
(Nomoto et al. 1995). This can turn UW CMa into a Type II-L, IIn, IIb,
Ib, or Ic of Supernova (Nomoto et al. 1995). The future evolution
of UW CMa is governed by large scale mass transfer and mass loss.
It may rapidly evolve into a luminous blue variable (LBV) and then evolve into
a Type II Supernova similar to LMC SN 1987A (Parthasarathy et al. 2006).
From the LBV phase it may alternatively evolve into a Wolf-Rayet binary
and end up again as a Type II Supernova.

\begin{table}
\tiny
\centering
 \caption{Photometric Solutions for Assumed Mass Ratios}
 \label{Phot_Sol}

\medskip

 \begin{tabular}{lcccccl}
 \hline
\noalign{\smallskip}
        & \multicolumn{5}{c}{$q=M_2/M_1$} & Parameter \\
\noalign{\smallskip}
 \cline{2-6}
\noalign{\smallskip}
\noalign{\smallskip}
 Parameters & $0.50^{\mathrm{b}}$ & 0.75 & 1.00 & 1.25 & 1.50 &  status \\
 \noalign{\smallskip}
 \hline
 \noalign{\medskip}

\hbox to 20mm{$i$ ($^{\circ}$) \dotfill}  &  $72.6$  & $71.3 \pm 0.6$ & $71.0 \pm 0.5$ & $71.2 \pm 06$ & $71.6 \pm 0.4$ & adjusted \\
\hbox to 20mm{$\Omega_1=\Omega_2$ \dotfill}   & $2.876$ &  $3.331 \pm 0.056$ & $3.750 \pm 0.059$ & $4.209 \pm 0.063$ & $4.526 \pm 0.066$ & adjusted \\
\hbox to 20mm{$\mu_1$ \dotfill}     & $0.999$ & $1.000 \pm 0.018$ & $0.994 \pm 0.022$ & $0.998 \pm 0.023$ & $0.997 \pm 0.024$ & adjusted \\
\hbox to 20mm{$\mu_2$ \dotfill}     & $0.997$ & $0.993 \pm 0.021$ & $0.999 \pm 0.020$ & $0.999 \pm 0.018$ & $1.000 \pm 0.019$ & adjusted \\
\hbox to 20mm{$T_1(K)$ \dotfill}    & $33750$ & $33750$         & $33750$ & $33750$ & $33750$  & adopted \\
\hbox to 20mm{$T_2(K)$ \dotfill}    & $32800$ & $33300 \pm 700$ & $33400 \pm 900$ & $33600 \pm 700$ & $33700 \pm 800$ & adjusted \\
\hbox to 20mm{$L_1/(L_1+L_2)^{\mathrm{a}}$ \dotfill} & $0.662$ & $0.570$ & $0.501$ & $0.451$ & $0.409$ & computed \\
\hbox to 20mm{$L_2/(L_1+L_2)^{\mathrm{a}}$ \dotfill} & $0.338$ & $0.430$ & $0.499$ & $0.549$ & $0.591$ & computed \\
\hbox to 20mm{$F_1$ \dotfill}      & 1.0     & 1.0     & 1.0    & 1.0     & 1.0  & adopted \\
\hbox to 20mm{$F_2$ \dotfill}      & 1.0     & 1.0     & 1.0    & 1.0     & 1.0  & adopted \\
\hbox to 20mm{$\beta_1$ \dotfill}  & 0.25    & 0.25    & 0.25   & 0.25    & 0.25 & adopted \\
\hbox to 20mm{$\beta_2$ \dotfill}  & 0.25    & 0.25    & 0.25   & 0.25    & 0.25 & adopted \\
\hbox to 20mm{$A_1$ \dotfill}      & 1.0     & 1.0     & 1.0    & 1.0     & 1.0  & adopted \\
\hbox to 20mm{$A_2$ \dotfill}      & 1.0     & 1.0     & 1.0    & 1.0     & 1.0  & adopted \\
\hbox to 20mm{$x_1$ \dotfill}      & -0.188  & -0.188  & -0.188 & -0.188  & -0.188  & adopted  \\
\hbox to 20mm{$y_1$ \dotfill}      & 0.719   & 0.719   & 0.719  & 0.719   & - 0.719   & adopted  \\
\hbox to 20mm{$x_2$ \dotfill}      & -0.141  & -0.141  & -0.141 & -0.141  & - -0.141  & adopted  \\
\hbox to 20mm{$y_2$ \dotfill}      & 0.746   & 0.746   & 0.746  & 0.746   & 0.746   & adopted  \\
\hbox to 20mm{$e$ \dotfill}        & 0. & 0. & 0. & 0. & 0. & adopted \\
\hbox to 20mm{$\omega$ \dotfill}    & 0. & 0. & 0. & 0. & 0. & adopted \\
\hbox to 20mm{$l_3$ \dotfill}      & 0.      & 0.      & 0.     & 0.      & 0.   & adopted \\
  \\
\hbox to 20mm{$\chi^2$ \dotfill}   & 250 & 235 & 230 & 235 & 232  & computed \\

\noalign{\smallskip}
 \hline
 \noalign{\smallskip}
 Relative radii (R/a) & & \\
 \noalign{\smallskip}
 \hline
 $r_1(pole)$  & $0.4143$ & $0.3802 \pm 0.0076$ & $0.3561 \pm 0.0071$ & $0.3376 \pm 0.0067$ & $0.3227 \pm 0.0064$ \\
 $r_1(point)$ & $0.5707$ & $0.5295 \pm 0.0061$ & $0.5000 \pm 0.0587$ & $0.4770 \pm 0.0568$ & $0.4583 \pm 0.0551$ \\
 $r_1(side)$  & $0.4399$ & $0.4009 \pm 0.0093$ & $0.3740 \pm 0.0086$ & $0.3537 \pm 0.0081$ & $0.3376 \pm 0.0077$ \\
 $r_1(back)$  & $0.4679$ & $0.4308 \pm 0.0125$ & $0.4050 \pm 0.0119$ & $0.3853 \pm 0.0114$ & $0.3696 \pm 0.0111$ \\

 $r_2(pole)$  & $0.2998$ & $0.3323 \pm 0.0066$ & $0.3561 \pm 0.0071$ &$0.3748 \pm 0.0075$ & $0.3902 \pm 0.0078$ \\
 $r_2(point)$ & $0.4292$ & $0.4705 \pm 0.0563$ & $0.5000 \pm 0.0587$ &$0.5229 \pm 0.0605$ & $0.5415 \pm 0.0619$ \\
 $r_2(side)$  & $0.3129$ & $0.3480 \pm 0.0080$ & $0.3740 \pm 0.0086$ &$0.3949 \pm 0.0092$ & $0.4122 \pm 0.0097$ \\
 $r_2(back)$  & $0.3454$ & $0.3797 \pm 0.0113$ & $0.4050 \pm 0.0119$ &$0.4250 \pm 0.0124$ & $0.4415 \pm 0.0128$ \\

 \noalign{\medskip}
 \hline
 \end{tabular}

\begin{list}{}{}
\item[$^{\mathrm{a}}$] $L_1,L_2$ - relative monochromatic luminosities of
the stars

\item[$^{\mathrm{b}}$] For q=0.5 the confidence intervals of the adjustable
parameters are not listed as the minimal deviations
exceed the critical value $\chi2$ at the confidence
level of $1\%$
\end{list}

\end{table}

\newpage


\begin{table}
\centering
 \caption{Absolute Parameters of UW CMa}
 \label{Abs_Par}

\medskip

 \begin{tabular}{lccccc}
 \hline
\noalign{\smallskip}
        & \multicolumn{5}{c}{$q=M_2/M_1$} \\
\noalign{\smallskip}
 \cline{2-6}
\noalign{\smallskip}
\noalign{\smallskip}
 {\hspace{0.5cm} Parameters} & 0.50 & 0.75 & 1.00 & 1.25 & 1.50  \\
 \noalign{\smallskip}
 \hline
 \noalign{\medskip}


\hbox to 25mm{$M_1 (M_{\odot})$ \dotfill}      & 106.9  &  44.0  &  24.4  & 15.8  &  11.2 \\
\hbox to 25mm{$M_2 (M_{\odot})$ \dotfill}     &  53.5  &  33.0  &  24.4  & 19.7  &  16.8 \\
\hbox to 25mm{$a(R_{\odot})$ \dotfill}        &  61.4  &  48.1  &  41.3  & 37.1  &  34.3 \\
\hbox to 25mm{$R_1 (R_{\odot})$ \dotfill}     &  27.2  &  19.5  &  15.7  & 13.4  &  11.8 \\
\hbox to 25mm{$R_2 (R_{\odot})$ \dotfill}     &  19.7  &  17.1  &  15.7  & 14.8  &  14.3 \\
\hbox to 25mm{$T_1 (K)$ \dotfill}            & 33750  & 33750  & 33750  & 33750  & 33750 \\
\hbox to 25mm{$T_2 (K)$ \dotfill}            & 32800  & 33300  & 33400  & 33600  & 33700 \\
\hbox to 25mm{$L_1 (10^5 L_{\odot})$ \dotfill} &  8.7   &   4.5  &   2.9  & 2.1  &   1.7 \\
\hbox to 25mm{$L_2 (10^5 L_{\odot})$ \dotfill} &  4.1   &   3.3  &   2.9  & 2.6  &   2.4 \\

 \noalign{\medskip}
 \hline
 \end{tabular}
 \end{table}

\section*{Acknowledgments}
The authors thank Prof  N. Samus and Dr I. Antokhin for useful
discussions. We also thank Prof S. Conti and the anonymous referee
for the constructive review of our paper. E. A.  acknowledges support
by the Russian Foundation for Basic Research through the
grant 08-02-01220.


\bibliographystyle{elsarticle-harv}
\bibliography{<your-bib-database>}



\end{document}